\documentstyle[aps,twocolumn,prb,epsf]{revtex}

\begin{document}
\draft

\twocolumn[\hsize\textwidth\columnwidth\hsize\csname @twocolumnfalse\endcsname

\title{Insulator-to-metal crossover induced by local spin fluctuations 
in strongly correlated systems}

\author{Luis Craco}

\address{Instituto de F\'{\i}sica ``Gleb Wataghin'' - UNICAMP, 
C.P. 6165, 13083-970 Campinas - SP, Brazil }

\date{\today}

\maketitle

\widetext
\begin{abstract}
\noindent
We study the simplified Hubbard (SH) model in the presence 
of a transverse field in the infinite dimension limit. 
The relevant one-particle Green's functions of the model are 
obtained by means a perturbative treatment of the hopping and of 
the transverse field around the atomic limit. We consider an analytical 
solution for the impurity problem. It is shown that this solution is 
very accurate in describing the spectral properties of the 
heavy-particles of the SH for intermediate and strong values of the 
on-site Coulomb interaction $U$. We find that for large values of $U$ 
an insulator-metal transition takes place as a function of the transverse 
field. We analyze the metallic phase through the behavior of the density
of states and of the optical conductivity and static resistivity. Our results
for the latter quantity agree with what is observed in experiments on 
$Bi_2Sr_2CuO_y$.  
\end{abstract}

]

\narrowtext
\section{Introduction}

In the last few years, several experimental works have considered the 
problem of the magnetic response in strongly correlated electronic systems.  
For example, the spin blockade problem in quantum dots in magnetic 
fields has been considered by different groups~\cite{qd,qd1}.  
The far-infrared transmission in thin films of $YBaCu_3O_7$ has been 
measured by Drew {\it et al}~\cite{Drew} and Lihn {\it et al.}~\cite{Lihn}, 
and a magneto-optical study of the magnetization for this 
compound was performed by Uspenskaya {\it et al.}~\cite{Uspen}.
The study of the magnetic properties of quantum dots~\cite{qd1} 
as well as high-$T_c$ thin films~\cite{BBKW,Boe,Ando} in presence of 
{\it transverse} magnetic fields has been also of great interest.

An important problem related with high-$T_c$ compounds is to understand 
the normal state properties of the in-plane $\rho_{ab}$ and and 
out-of-plane $\rho_c$ dc-resistivity for different values of the doping 
concentration~\cite{Ande}. 
It is well known that the easiest way to suppress the superconductivity 
at low temperatures without deliberate chemical substitution is to apply a 
high (pulsed) magnetic field. In such experiments the field is usually applied 
parallel to the $c-$axis to most effectively suppress the superconducting 
state, so that the magnetic field is normally applied perpendicular to the 
$CuO_2$ basal plane where the ordered spins are primarily 
aligned~\cite{Imada,Vak}. This alignment is usually originated by spin-orbit 
interaction and it has been observed on different alloys~\cite{dif-mat}. 

On the other hand, extensive experimental investigations
show that the normal state properties of high-$T_c$
superconductors are not explained in terms of the Fermi-liquid (FL) 
theory~\cite{Ande,Batl}. Deviations from the normal FL behavior
were observed also in the normal state of different heavy-fermion 
compounds~\cite{heavy} as well as in quasi-one-dimensional 
materials~\cite{qod}.
This unexpected  behavior increased the interest in studying physical 
models that clearly present non-Fermi liquid (NFL) properties.
For example, Si {\it et al.}~\cite{Si} introduced a spinless two-band model 
to describe the effect of interactions in a band insulating system.
By solving the model numerically in the limit of infinite spatial dimensions 
$(d \rightarrow \infty)$ they showed that it exhibits NFL properties. 
In addition, Consiglio {\it at al.}~\cite{Consiglio} 
have shown that the main features of the optical conductivity of the 
Kondo alloy $Y_{1-x}U_xPd_3$ are well taken into account by 
the simplified Periodic Anderson model. In the last years, attention has also
been given to the simplified Hubbard (SH)~\cite{BM,Janis,Si-FK,Pedro,Pavol}. 
This is a modification of the Hubbard model, where electrons with one 
particular spin orientation do not hop in the lattice.  
This is one of the few models which possess exact solution 
in the limit of high dimensions~\cite{BM}, and it shows a metal-insulator 
transition both as a function of the on-site Coulomb interaction and of 
the doping. 

Of particular importance here is the work of 
Brandt and Urbanek~\cite{BU}, where the spectral properties of the 
heavy-particles, the electrons that do not hop, are discussed in details. 
In addition, some years ago the SH model in a magnetic field has been 
studied by van Dongen and Leinung~\cite{vDL}. They consider the problem of
the metal-insulator transition as a function of the Zeeman ($z$-direction)
field. As expected, at large fields the system is a fully polarized 
ferromagnet, however in the non-saturated phase a metal-insulator (MI) 
transition takes place as a function of the field and the local one-site 
Coulomb interaction $(U)$. They show that magnetic field slightly reduces 
the critical value of the Coulomb interaction, so that the MI transition for 
non-zero field occurs at a critical $U$ smaller then the one at zero field.

It is worth noticing that, despite of the great interest in strongly 
correlated electron systems, to the best of our knowledge there is no 
investigation of the effects of transverse  
fields in models which explicitly shows both broken spin symmetry and 
non-Fermi liquid properties. In this paper, we address precisely to this 
issue: the study of the SH model in the presence of a transverse field 
in the the high dimension (D) limit. This limit, introduced 
originally by Metzner and Vollhardt~\cite{MV}, has been shown to be a very 
good starting point in the study of several physical systems~\cite{Georges}. 
We study the SH  model because it has by construction broken spin symmetry, 
which is closely connected to the different dynamics of the two types of 
electrons. Here, we discuss the formal solution of this model in a 
transverse field, where the so called {\it static approximation}~\cite{LM2} 
is employed to solve the related single-site problem. Comparing with the 
results of Ref. [23], 
we show that the {\it static approximation} gives a very accurate 
description of the $T=0$ physical properties of the heavy-particles 
for on-site Coulomb interactions above the metal-insulator transition.
Next, we study the effect of the transverse field on the spectral 
and the optical properties of the SH model.

\section{Model and perturbation method}
\label{sec:pert}

In the usual notation, the SH model in a transverse field 
is described by the Hamiltonian

\begin{eqnarray}
\label{eq:model}
\nonumber
H & = & \sum_{\bf k} \epsilon_{\bf k} 
c_{\bf k \uparrow}^\dagger c_{\bf k \uparrow}  
+ U \sum_{i} n_{i\uparrow} n_{i\downarrow} 
+ (E-\mu) \sum_{i\sigma}n_{i \sigma}  \\ & + &
\sum_{i} t  (c_{i \uparrow}^\dagger c_{i \downarrow} + h. c.)
\;, 
\end{eqnarray}  
where $\epsilon_{\bf k}$ is the dispersion relation of the conduction 
\mbox{$(\uparrow$-electrons)}, and the  
\mbox{$\downarrow$-electrons} do not hop in this model. 
$E$ is the energy level of the two particles, which are 
coupled through the correlation $U$ and a transverse field $t$. 
This field is usually a magnetic field along the $x$-direction, but 
similar type of effect could have a different origin. One example
would be a correlated hopping with spin flip originated in the spin orbit
interaction, and this type of term has been already employed~\cite{SO} 
in the study of other strongly correlated electron systems. 

In this work we are interested in the two relevant one-particle 
Green's functions of Eq.~(\ref{eq:model}). The temperature-dependent 
one-particle Green's function for both up and down electrons are obtained,  
following the approach recently introduced to study the formal solution 
of a spinless two-band model~\cite{p2}, by means a perturbative treatment 
around the atomic limit of the hopping and the transverse field 
terms. To apply this method, we first consider the  exact solution of the 
unperturbed Hamiltonian, given by the last two terms of the {\it first} 
line of Eq.~(\ref{eq:model}). Next we solve the $t=0$ limit of 
Eq.~(\ref{eq:model}) by means of a tight-binding treatment around the 
atomic limit of the conduction electrons~\cite{LM1}. 
The formal solution of the complete Hamiltonian is 
then obtained by performing a perturbative treatment on the 
hybridization term~\cite{p2}. Following this procedure and considering 
the high dimension limit, it is straightforward to show that 

\begin{equation}
\label{eq:giiup}
G_{ii \uparrow}  (i\omega_n) =\frac{1}{N}\sum_{\bf k}  \frac{1}{ 
[ {\cal G}_\uparrow (i\omega_n) ]^{-1} 
-\epsilon_{\bf k} - t^2 {\cal G}_\downarrow (i\omega_n) }
\end{equation}
and
\begin{equation}
\label{eq:giidown}
G_{ii \downarrow}  (i\omega_n) = {\cal G}_\downarrow (i\omega_n) [ 1 +  
{\cal G}_\downarrow (i\omega_n) t^2 G_{ii \uparrow} (i\omega_n) ] \;,
\end{equation}
where ${\cal G}_{\uparrow} (i\omega_n)$
and ${\cal G}_{\downarrow} (i\omega_n)$ are the  
irreducible one-particle Green's functions for the 
conduction electrons and for the heavy-particles, respectively. 
The former is irreducible in the sense that the contributing diagrams 
can not be divided in two pieces by cutting a single hopping line, 
while the second is irreducible with respect to the cutting 
a $t^2 {\bar g}_{\bf k \uparrow} (i\omega_n)$ line~\cite{Consiglio,p2}, 
where ${\bar g}_{\bf k \uparrow} (i\omega_n)$ is the solution of 
Eq.~(\ref{eq:model}) in the limit of $t=0$. From Eq.~(\ref{eq:giiup})
it becomes clear that the transverse field $t$ acts as an hybridization
term, mixing the the singe-site one-particle excitation of the 
$\uparrow$-electrons with the $\downarrow$-ones. One should notice that
in the case of complete Hubbard model a similar equation for the  
$\downarrow$-electrons is obtained by exchanging the spin directions.
The main difference between Eqs.~(\ref{eq:giiup}-\ref{eq:giidown}) and 
those of Ref. [29] is that here they describe the formal solution
of one band model with hybridization in the spin sector, while there
we considered a two band model with interaction and hybridization
in the charge degrees of freedom of fully polarized orbitals~\cite{nsf}. 

The irreducible propagators in Eqs.~(\ref{eq:giiup}-\ref{eq:giidown})
can be written in terms of the single-site one-particle Green's function 
and the dynamical mean-field ${\cal A}_\sigma (i\omega_n)$, which 
connects a single-site with the electron bath, through the 
relation~\cite{LM2}

\begin{equation}
\label{eq:aa}
\frac{1}{ {\cal G}_\sigma (i\omega_n) } = 
\frac{1}{ G_{ii \sigma} (i\omega_n) } + {\cal A}_\sigma (i\omega_n) \;.
\end{equation}

It is important to notice that once we turn on the transverse field, the 
solution of the conduction electron local problem is not trivial
any longer. This is because the transverse field term hybridizes 
locally the up and down electrons, so that the spin-down electrons
are not any longer frozen. Once we provide dynamics to the heavy-electrons,
the local spin fluctuation problem also holds for the conduction electrons.

Now we proceed to analyze the solution of the single-site problem of both
electrons from the point of view of the perturbation around the atomic 
limit~\cite{LM2,p2}. This method provides a direct way of solving the 
local problem by means of a perturbative expansion in the local mean-field. 
This approach has been used in Ref. [27] to obtain the exact solution 
for the conduction electrons of the SH model as well as to study the problem 
of the spin fluctuation in the Hubbard model. For the latter case the    
{\it static approximation} has been introduced. Here, we will employ this 
approximation to account for the local spin fluctuations induced by the 
transverse field.

In our perturbation approach to the single-site problem we start with the 
unperturbed local Green's function 
$-\langle \hat T c_{\sigma}^{}(\tau) c_{\sigma}^{\dagger}(0) \rangle_0$,
each order in perturbation theory introduces a product of the type
${\cal A}_{\sigma_1} (\tau_1 - \tau^{\prime}_1) c_{\sigma}(\tau_1) c_{\sigma}^{\dagger}(\tau^{\prime}_1)$ so that in general one has to calculate averages of the form $\langle c_{\sigma_1}^{}(\tau_1) c_{\sigma_1}^{\dagger}(\tau^{\prime}_1)c_{\sigma_2}^{}(\tau_2) c_{\sigma_2}^{\dagger}(\tau^{\prime}_2) \rangle_0$. 
To calculate this averages we rewrite the fermions operators in terms of 
the Hubbard operators~\cite{LM1}, and utilize the standard algebra for the
latter. This allows us to perform all possible direct contractions, 
in the sense of Wick's theorem. We have applied this procedure in 
Refs. [27,30] to evaluate explicitly a four-operator average 
that appears in the one-loop approximation. In addition, in Ref. [27]
the {\it static approximation} for the single-site one-particle Green's 
function has been obtained by neglecting all terms that involve non-zero
bosonic frequencies originated by local contractions between boson like
Hubbard operators. Since the averages we need to
consider here are exactly the same as in the case of the Hubbard model 
we refer the reader to Refs. [27,30] for their calculation.

In terms of the {\it static approximation} the single-site one-particle 
Green's function is given by

\begin{equation}
\label{eq:static}
G_{ii \sigma} (i\omega_n) = \bar G_{ii \sigma} (i\omega_n) +
\Delta_{\sigma} (i\omega_n) \;,
\end{equation}
where 
\begin{equation}
\label{GFK}
\bar G_{ii \sigma} (i\omega_n) =
(1-\langle n_{\bar \sigma} \rangle) \bar g_{0 \sigma} (i\omega_n) 
+ \langle n_{\bar \sigma} \rangle \bar g_{\bar \sigma 2} (i\omega_n)
\end{equation}
is the well known exact solution for the conduction electrons of 
the SH model (the $t=0$ limit of Eq.~(\ref{eq:model})) and
\begin{eqnarray}
\label{delta}
\Delta_{\sigma}(i\omega_n) & = &
\left(g_{0\sigma} - g_{\bar \sigma 2} \right) {\cal A}_{\bar \sigma}
\Bigl[ \frac{(1-\langle n_{\bar \sigma} \rangle) \bar g_{0 \sigma}}
{1-  {\cal A}_{\sigma} g_{0 \sigma} - 
{\cal A}_{\bar \sigma} (g_{0\sigma} + g_{\bar \sigma 2})}
\nonumber \\ 
& - & \frac{ \langle n_{\bar \sigma} \rangle  \bar g_{\bar \sigma 2}}
{1-  {\cal A}_{\sigma} g_{\bar \sigma 2} - 
{\cal A}_{\bar \sigma} (g_{0\sigma} + g_{\bar \sigma 2})}
\Bigr]  \;.
\end{eqnarray}
$g_{0\sigma} (i\omega_n)$ and  
$g_{\bar \sigma 2} (i\omega_n)$ in Eqs.~(\ref{GFK}-\ref{delta})
are the fermionic Green's 
functions of Hubbard operators~\cite{LM1}, while 
${\bar g_{0 \sigma}} (i\omega_n)$ and ${\bar g_{\bar \sigma 2}} (i\omega_n)$ 
describe the single-site one-particle excitations renormalized 
by the dynamical mean-field~\cite{barg}. Note that even for the lowest order  
in the dynamical mean-field, $\Delta_{\sigma}(i\omega_n)$ is obtained from 
local contractions between {\it one-band} operators with different spin 
orientations. This means that $\Delta_{\sigma}(i\omega_n) =0$ for systems 
where local spin fluctuations are absent~\cite{nsf}.  

\section{Results}
\label{sec:results}

\begin{figure}[htb]
\epsfxsize=3.3in
\epsffile{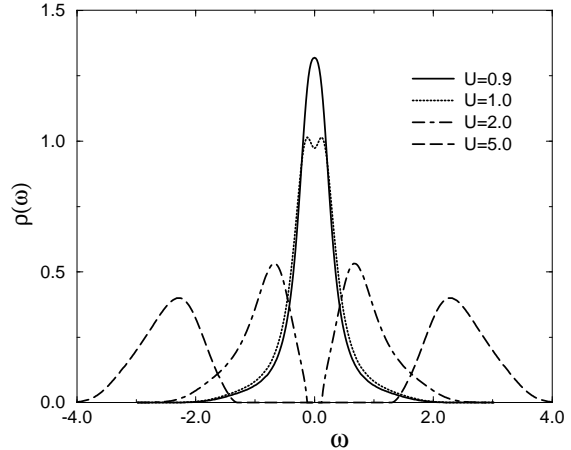}
\caption{Zero temperature densities of states for the 
heavy-electrons of the simplified Hubbard model at t=0 for 
different values of $U$.} 
\label{fig1}
\end{figure}

Let us consider first the $t=0$ limit of Eq.~(\ref{eq:model}).
${\cal A}_{\downarrow} (i\omega_n) = 0$ in this limit, because 
${\cal A}_{\downarrow}$ is proportional to the hybridization
term. This can be easily seen from Eqs.~(\ref{eq:giidown}) 
and~(\ref{eq:aa}). Therefore, in this limit the single-site 
one-particle Green's function of the heavy-electrons 
$G_{ii \downarrow} (i\omega_n)$ is only a function of 
the dynamical mean-field of the conduction 
electrons, see Eqs.~(\ref{eq:static}-\ref{delta}). 
In Fig.~\ref{fig1} the density of states (DOS) of the  
heavy-electrons is shown for $t=0$ and different values of 
$U$~\cite{calc}. For $U=0.9$ we obtain a sharp peak around the 
Fermi level $(\omega=0)$. Increasing $U$, the height of the central 
peak is reduced, and a small tendency to form a gap is observed
for $U=1$. From the comparison between 
our results with those of Ref. [23] it becomes clear that 
the {\it static approximation} is able to recover the main features of the 
spectral properties of the heavy-particles in the metallic phase of 
the SH model. Indeed, in the regime of strong on-site interaction $U$ we 
observe a very good agreement between our results and those 
obtained by Brandt {\it et al.}~\cite{BU}. 
Hence, one can conclude that the {\it static approximation} 
is very accurate in describing the $\downarrow$-electrons properties 
of the SH model in the strong coupling limit. 

From this result it is possible to conclude that for the half-filled and 
symmetric SH model at $T=0$ all the terms that involve non-zero 
bosonic frequencies~\cite{BU,LM2,LM1} 
in the local Green's functions do not play an important 
role in the limit of large values of the Coulomb on-site interaction.
Indeed, Fig.~\ref{fig1} shows that the perturbation treatment around the
atomic limit is very accurate in the large $U$ limit. Note that, in our 
treatment for the corrections due to spin fluctuations (Eq.~(\ref{delta})) 
an infinite series of diagrams for the site-diagonal Green's function as 
well as for the irreducible propagators are considered. Furthermore, based 
on the results of Fig.~\ref{fig1} it becomes clear that our solution for 
the single site problem is a good approximation, and we do not need to 
consider any small parameter to justify Eq.~(\ref{delta}).

\begin{figure}[htb]
\epsfxsize=3.2in
\epsffile{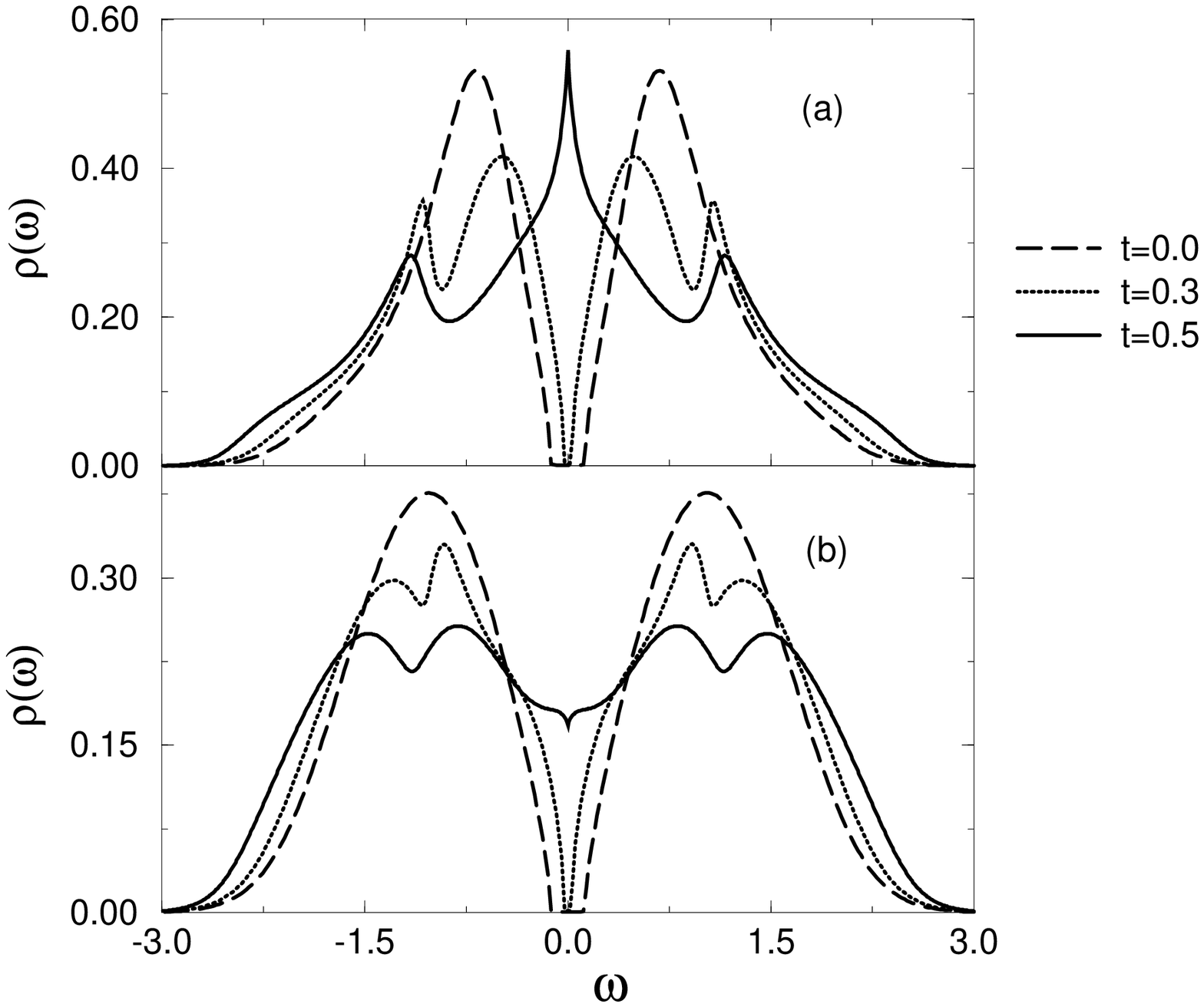}
\caption{Zero temperature densities of states for the SH model   
for $U=2$ and different values of $t$. (a) and (b) are the 
DOS of $\downarrow$ and $\uparrow$-electrons, respectively.} 
\label{fig2}
\end{figure}

As our method describes correctly the spectral properties of the SH model 
in the large $U$ limit, we shall employ this limit to study the spectral 
properties of the SH model in the presence of the transverse magnetic field,
expecting to obtain realistic results. 
In Fig.~\ref{fig2} the DOS of both types of electrons is shown for
$U=2$ and different values of the transverse field $t$. 
Let us consider once more the $t=0$ limit. In the dashed line 
of Fig.~\ref{fig2} one can clearly see that the gap size for both  
particles coincide. Hence, it is clear from this behavior that 
the DOS at the Fermi level vanishes at the same value of $U$
for both spin directions. Furthermore, the DOS of both particles 
shows very strong similarities. One can expect that these similarities 
will increase with $U$, and that both particles would behave almost in 
the same way at the half-filled case for $U\rightarrow\infty$.

According to our results in Fig.~\ref{fig2}, the charge Mott gap is 
strongly affected by the transverse field. In the case of $U=2$, this gap 
persists up to $t=0.3$, where the insulator-metal transition take place. 
It is instructive to comment that for both spin directions we observe  
spectral transfer from the high to low energies. 
The most interesting behavior of the metallic phase occurs for 
$t=0.5$, where a sharp peak around the Fermi level is observed in 
the DOS of the heavy particles. This peak is related  
with a self-consistent modification of the electron bath, 
where a partial decoupling of the local degrees of 
freedom take place~\cite{Si}. Note that, for both nonzero values of the
transverse field in Fig.~\ref{fig2} the two atomic-like poles of the 
$G_{ii\uparrow}$ remain almost in the same position, while the internal 
pole of the $G_{ii\downarrow}$ is shifted to zero-energy for $t=0.5$. 
This is because the transverse field strongly mixes the low energy
excitations of the two types of electrons by spin flip process.

\begin{figure}[htb]
\epsfxsize=3.2in
\epsffile{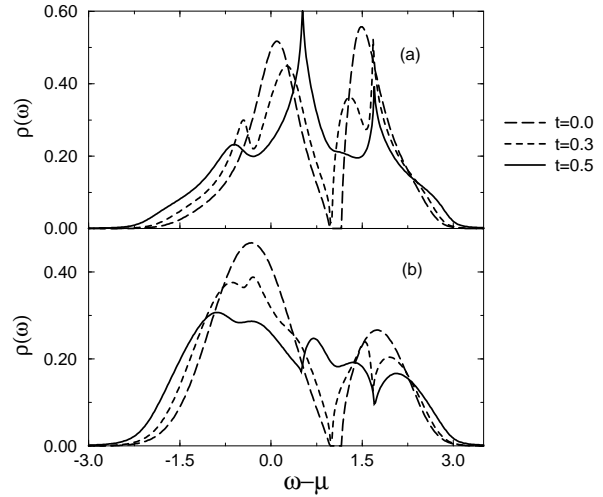}
\caption{Zero temperature densities of states for the SH model   
for $U=2$, $n=0.8$ and different values of $t$. (a) and (b) are the 
DOS of $\downarrow$ and $\uparrow$-electrons, respectively.} 
\label{fig3}
\end{figure}

To provide a complete description of the insulator-metal transition
of the SH model in a transverse field we study in Fig.~\ref{fig3}
the single-particle density of states away from half-filling, 
at a fixed $n=0.8$. For $t=0$ one can
see that our results for the DOS of the conduction electrons agrees with those
obtained by M\"oller {\it et al.}~\cite{Moller}. As in Fig.~\ref{fig2}
the gap in the DOS of both  particles coincide in this limit.
However, the upper Hubbard band of the heavy-electrons 
is higher than the one for the conduction electrons, and this behavior is
related with the atomic-like character of the heavy-particles at large 
$U$. Another interesting feature to be seen in Fig.~\ref{fig3} is the 
presence of a second sharp peak at high energies in the DOS of the 
$\downarrow$-electrons for non-zero $t$.
This peak reduces by increasing the transverse field. 
Finally, the DOS of the conduction electrons also shows interesting new 
structures, in particular for $t=0.5$. We believe that these structures 
are related with the asymmetric behavior in the DOS of the heavy-electrons 
together with the large hybridization between the two types of electrons.

To further confirm the insulator-metal transition we calculate
the optical conductivity for the half-filled case with the same
parameters employed in  Fig.~\ref{fig2}. 
It is known that in the infinite dimension limit the vertex corrections
drop out in the two particle equation, and the optical conductivity
assumes the simple form~\cite{Georges}

\begin{eqnarray}
\label{opt-cond}
\sigma(\omega) & = & \pi \sum_\sigma 
\int d\epsilon \rho_0 (\epsilon) 
\int d\omega^\prime A_\sigma(\epsilon,\omega^\prime)
A_\sigma(\epsilon,\omega^\prime + \omega) 
\nonumber \\ && \times 
\frac {[f(\omega^\prime) - f(\omega^\prime+\omega)}{\omega} \;,
\end{eqnarray}
where $\rho_0 (\epsilon)$ is the uncorrelated density of states 
of the conduction band, $f(\epsilon)$ is the Fermi function and 
$ A_\sigma(\epsilon,\omega^\prime)$ is the one-particle spectral 
density of the total conduction-electrons Green's function. 

\begin{figure}[htb]
\epsfxsize=3.3in
\epsffile{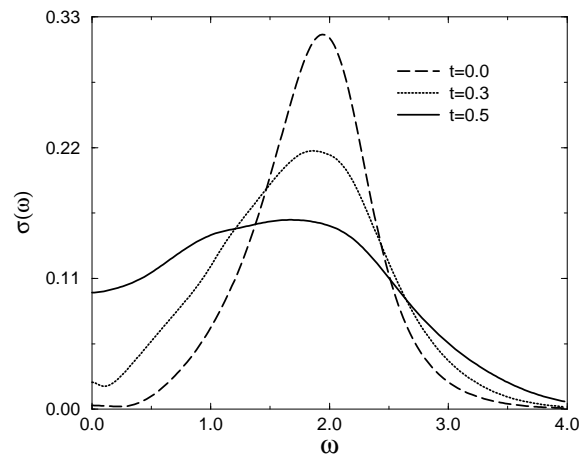}
\caption{Zero temperature optical conductivity for the SH model in the  
half-filled case for $U=2$,  and three different values of the applied 
transverse magnetic field.} 
\label{fig4}
\end{figure}

Our results for the optical conductivity are shown in Fig.~\ref{fig4}.
For $t=0$ we have a very good agreement with those of Ref. [35]. 
As one can see, the system is insulator with a large gap in the optical 
conductivity and
the maximum value of $\sigma(\omega)$ coincides with the value of the Hubbard 
interaction. Once we turn on the transverse field, we observe a spectral 
transfer from high to low energies, as in the DOS. The spectral 
transfer closes the gap and increases the conductivity at low frequencies.
As we expect the system is a incoherent metal (no Drude peak feature is 
found) mainly because the transverse field is not able to completely restore  
the dynamics of the heavy-particles.

\begin{figure}[htb]
\epsfxsize=3.3in
\epsffile{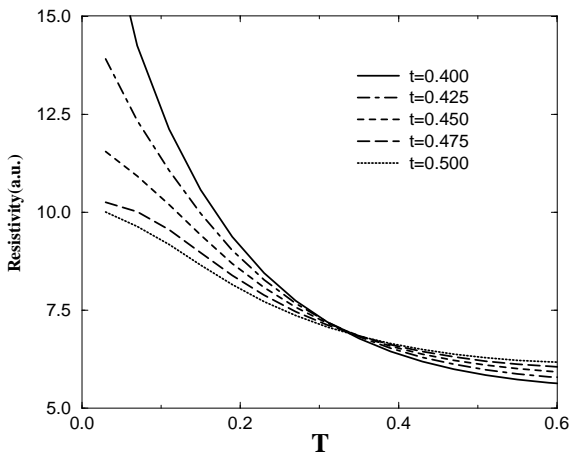}
\caption{Temperature dependence of the static resistivity in the half-filled
SH model for $U=2$ and different values of $t$.} 
\label{fig5}
\end{figure}

Now we turn our attention to Fig.~\ref{fig5} where we show our results
for the dc-resistivity at half-filling as a function of the temperature. 
In $d=\infty$ this quantity is given by the inverse $(1/\sigma(0))$ of the 
static limit $(\omega=0)$ of Eq.~(\ref{opt-cond})

\begin{equation}
\label{st-res}
\sigma(0)=\frac{\pi}{T} \int d\epsilon \rho_0 (\epsilon) 
\int d\omega^\prime A_{\uparrow}^{2}(\epsilon,\omega^\prime) 
f(\omega^\prime)[1-f(\omega^\prime)] \;,
\end{equation}
where $T$ is the temperature.

At $t=0$ and $U=2$ the dc-resistivity shows a semiconducting-like behavior
with temperature. Above $T=0.6$, not shown, it starts to 
increase with $T$ as in a metal. In accordance with our previous results 
for the insulator-metal transition the static resistivity decreases
as function of the transverse field. One can clearly see in Fig.~\ref{fig5}
the presence of an isosbectic point, where all the curves crosses at 
the same temperature. It is important to mention that a similar 
behavior for the static resistivity as the one we obtain for $0.45<t<0.5$
has been observed in the normal state of $Bi_2Sr_2CuO_y$.

In Ref. [8] the in-plane $(\rho_{ab})$ and out-of-plane 
$(\rho_{c})$ dc-resistivity of $Bi_2Sr_2CuO_y$
have been measured for different values of the 
magnetic field. The experiment shows completely different behavior 
for the in-plane and out-of-plane measurements: $\rho_{ab}$ increases 
to a certain saturation value, while $\rho_{c} $ strongly decreases (by 
a factor $10^4$) for the same values of the magnetic field. 
One can understand this contrasting behavior taking into account our
results and those of Ref. [24]. As mentioned before, it was shown
in Ref. [24] that the critical $U$ for the metal-insulator 
crossover is reduced by the effect of a $z$-component magnetic field. 
Thus for the same value of $U$ the system loses its metallic phase 
as a function of the field and the resistivity increases with 
the Zeeman field. On the other hand,  we show here that a field applied 
along to the $x$-direction (in the spin space) drives and insulator to a
metal transition in the system. Since both theoretical results agree
with was observed in the measurements of {\it Ando et al.}, one can conclude
that the direction along which the field is applied is essential in 
systems with broken spin symmetry. 

\section{Conclusions}
\label{sec:concl}

We report the results of the first systematic study of a strongly correlated 
electron system in transverse fields, undertaken to understand the interplay 
between on-site correlations and local spin fluctuations. To make contact with
experiment we consider the simplified Hubbard (SH) model, a model Hamiltonian 
with broken spin symmetry. The relevant one-particle Green's functions for 
this model in the presence of a transverse field $t$ are obtained by means 
of a perturbative treatment of the hopping and the field around the atomic 
limit. We employ the {\it static approximation} to study the insulator-metal 
transition as a function of $t$. At intermediate values of the field the 
metallic phase shows very interesting features. In this regime we observe the 
presence of sharp peaks in the spectral function of the heavy-electrons 
at the Fermi level. We show that the {\it static approximation}
is very accurate in describing the dynamics of the heavy-particles 
in intermediate and strong values of $U$. 

We have also studied the optical properties of the SH model as a function of 
the transverse field. Our results for the optical conductivity and
static resistivity confirm the insulator-metal crossover. 
We found good qualitative agreement between our results for the 
latter quantity and those observed in the out-of-plane resistivity 
of $Bi_2Sr_2CuO_y$~\cite{Ando}. In both theoretical and experimental 
investigations the resistivity decreases as a function of the  
magnetic field. In addition, we observe in our results the presence of an 
{\it isosbectic} point, where all curves cross at the same temperature.
This crossing point seems to appear also in the experiment, although it
could not be clearly observed because the measurements were done for 
temperatures slightly below the occurrence of the crossing.  

To further our understanding about the semiconducting behavior along 
to the $c$-axis in cuprates~\cite{Iye}, we should mention that in 
the non-magnetic phase the SH model has an 
intrinsic disorder built into it. Such disorder is responsible for the  
non-Fermi liquid properties of the conduction electrons, since the 
heavy-electrons act as randomly distributed scattering centers. For large 
values of disorder ($U$ in the present model) the motion of the 
conduction electrons is totally suppressed and the system behaves as 
a semiconductor, as shown in Fig.~\ref{fig5} for small values of $t$.
According to our results, the semincoducting behavior can be partially 
suppressed by enhancing the local spin degrees of freedom through a
transverse field. Following this scenario and the measurements
of the $dc$-resistivity in $Bi_2Sr_2CuO_y$ one can suggest that
the low temperature semiconducting behavior in cuprates along 
the $c$-direction might result from conventional random disorder.

Finally, as a possible extension 
of this work we should consider in the future the possibility of extending 
our approach to the case of the one-band Hubbard model.
In this case it would be interesting to analyze the effect of the 
transverse field on the well known Fermi-liquid properties of this model.
Furthermore, inspired by the results of Fig.~\ref{fig1} 
it would be interesting to analyze whether our perturbation method 
for the local problem is also accurate in describing the single particle
properties of the Hubbard model in the large $U$ limit. Unfortunately,  
apparently there are no exact results for this model 
at $T=0$ in $d=\infty$, but one can follow the spirit of Ref. [29] 
and compare our results for the Matsubara Green's functions with those of 
finite temperature quantum Monte Carlo simulations~\cite{Georges}.  
We believe that the {\it static approximation} (Eq.~(\ref{eq:static})) 
is an effective approximation for the Hubbard model if we consider that 
the {\it local} spin fluctuation are irrelevant in the strong coupling limit.

\acknowledgments
The author wishes to acknowledge M. Foglio, C. Kuebert and P. Farinas 
for useful discussions and comments. This work was supported by the 
Funda\c c\~ao de Amparo \`a Pesquisa do Estado de S\~ao Paulo (FAPESP).


\begin{references}

\bibitem{qd} D. Goldhaber-Gordon {\it et al.} Nature {\bf 391}, 156 (1998);
H. Imamura {\it et al.} Physica {\bf B 256-258}, 194 (1998);
M. Pustilnik {\it et al.} preprint con-mat/9908004.

\bibitem{qd1} S. M. Cronenwett {\it et al.}, Science {\bf 281}, 540 (1998). 

\bibitem{Drew} H. D. Drew {\it et al.} J. Phys. Cond. Mat. {\bf 8}, 
10037 (1996). 

\bibitem{Lihn} H. T. S. Lihn {\it et al.} Phys. Rev. Lett. {\bf 76},
3810 (1996).

\bibitem{Uspen} L. S. Uspenskaya {\it et al.} Phys. Rev. {\bf B 56},
11979 (1997). 

\bibitem{BBKW} J. Buan {\it et al.} Phys. Rev. Lett. {\bf 72},
2632 (1994); E. H. Brandt Phys. Rev. Lett. {\bf 74}, 3025 (1995); 
A. V. Kuznetsov {\it et al.} Phys. Rev {\bf B 52} 9637 (1995); 
M. Wurlitzer {\it et al.} Phys. Rev. {\bf B 55}, 11816 (1997).

\bibitem{Boe} G. S. Boebinger {\it et al.} Phys. Rev. Lett. 
{\bf 77}, 5417 (1996).

\bibitem{Ando} Y. Ando {\it et al.} Phys. Rev. Lett. {\bf 77}, 2065 (1996).

\bibitem{Ande} P. W. Anderson in ``The Theory of Superconductivity 
in the High-$T_{c}$ cuprates'', Princeton University Press (1997).

\bibitem{Imada} M. Imada {\it et al.} Rev. Mod. Phys. {\bf 70},
1039 (1999).

\bibitem{Vak} D. Vaknin {\it et al.} Phys. Rev. Lett. {\bf 58},
2802 (1987); E. Brecht {\it et al.} Phys. Rev. {\bf B 52}, 9601 (1995).

\bibitem{dif-mat} K. Terakura {\it et al.} Phys. Rev. {\bf B 30}, 
4734 (1984); W. Bao {\it et al.} Phys. Rev. Lett {\bf 71}, 766 (1993); 
G. Blumberg {\it et al.} Phys. Rev. Lett. {\bf 80}, 564 (1998).

\bibitem{Batl}  B. Batlogg, in ``The Los Alamos Symposium'' eds. R. Coffey 
{\it et al.} Addison Wesley Publ. (1991).

\bibitem{heavy} C. L. Seaman {\it et al.} Phys. Rev. Lett. {\bf 67}, 
2882 (1991); H. V. L\"ohneysen {\it et al.} Phys. Rev. Lett. 
{\bf 72}, 3262 (1994); L. Degiorgi {\it et al.} Phys. Rev. {\bf B 52}, 
42 (1995); M. B.  Maple {\it et al.} Physica {\bf B 223-224}, 
447 (1996); F. Bommeli {\it et. al.} Phys. Rev. {\bf B 56} 10001 (1997).

\bibitem{qod} J. Voit Rep. Prog. Phys. {\bf 58}, 977 (1995).

\bibitem{Si} Q. Si {\it et al.}, Phys. Rev. Lett. {\bf 72}, 2761 (1994).

\bibitem{Consiglio} R. Consiglio and M. A. Gusm\~ao, Phys. Rev. {\bf B 55},
6825 (1997).

\bibitem{BM} U. Brandt and C. Mielsch, Z. Phys. {\bf B 75}, 365 (1989).

\bibitem{Janis} V. Jani$\breve{s}$, Z. Phys. {\bf B 83}, 227 (1991).

\bibitem{Si-FK} Q. Si {it et al.}, Phys. Rev. {\bf B 46}, 1261 (1992).

\bibitem{Pedro} P. de Vries {\it et al.} Z. Phys. {\bf B 92}, 353 (1993).

\bibitem{Pavol} P. Farka$\breve{s}$ovsk\'y, Phys. Rev. {\bf B 51},
1 507 (1995).

\bibitem{BU} U. Brandt and M. P. Urbanek, Z. Phys. {\bf B 89}, 297
(1992).

\bibitem{vDL} P. G. J. van Dongen and C. Leinung, Annalen der Physik
{\bf 6}, 45 (1997).

\bibitem{MV}  W. Metzner and D. Vollhardt, Phys. Rev. Lett.  {\bf 62}, 
324 (1989); see also E. M\"uller-Hartmann, Z. Phys. {\bf B 74}, 507 (1989).

\bibitem{Georges} A. Georges {\it et al.}, Rev. Mod. Phys. {\bf 68}, 
13 (1996).

\bibitem{LM2} L. Craco and M. A. Gusm\~ao,  Phys. Rev. {\bf B 54}, 1629
(1996).

\bibitem{SO} N. E. Bonesteel {\it et al.}, Phys. Rev. Lett. {\bf 68}, 2684
(1992); T. Yildirim {\it et al.}, Phys. Rev. Lett. {\bf 73}, 2919 (1994). 

\bibitem{p2} L. Craco, Phys. Rev. {\bf B 59}, 14 837 (1999).

\bibitem{LM1} L. Craco and M. A. Gusm\~ao, Phys. Rev. {\bf B 52},
17135 (1995).

\bibitem{nsf} The spin degeneracy can be lifted by a strong magnetic field 
or by a very large Hund's-rule coupling, as in the case of
$Fe_3O_4$~\cite{Cullen}.

\bibitem{Cullen} J. R. Cullen and E. Callen, Phys. Rev. Lett. 
{\bf 26}, 236 (1971).

\bibitem{barg} More precisely, $\bar g_{0 \sigma} (i\omega_n) 
\equiv 1/(i\omega_n - E +\mu - {\cal A}_{\sigma}(i\omega_n))$ and 
$\bar g_{\bar \sigma 2} (i\omega_n) 
\equiv 1/(i\omega_n - E - U +\mu - {\cal A}_{\sigma}(i\omega_n))$. 

\bibitem{calc} The calculations were performed at the symmetric case with 
a Gaussian DOS. 

\bibitem{Moller} G. M\"oller {\it et al.} Phys. Rev. {\bf B 46}, 7427 (1992).

\bibitem{Iye} Y. Iye, in ``Physical Properties of High Temperature 
Superconductors $III$, edited by D. M. Ginsberg (World Scientific,
Singapore, 1991).

\end{references}
\end{document}